\documentstyle[12pt]{article}
\begin{document}
{\pagestyle{empty}
\rightline{TU-01/97}
\rightline{January 1997}
\rightline{~~~~~~~~~}
\vskip 1cm
\centerline{\Large \bf Parallel Transport in Gauge Theory}
\centerline{\Large \bf on $M_4 \times Z_2$ Geometry}
\vskip 1cm

\centerline{{Takesi Saito\footnote{E-mail address: 
             tsaito@jpnyitp.yukawa.kyoto-u.ac.jp}} and 
            {Kunihiko Uehara\footnote{E-mail address:
             uehara@tezukayama-u.ac.jp}}}
\vskip 1cm
\centerline{\it Department of Physics, Kwansei Gakuin University, 
Nishinomiya 662, Japan${}^1$}
\centerline{\it Department of Physics, Tezukayama University, 
Nara 631, Japan${}^2$}

\vskip 2cm

\centerline{\bf Abstract}
\vskip 0.2in

  We apply the gauge theory on $M_4\times Z_2$ geometry previously 
proposed by Konisi and Saito to the Weinberg-Salam model for 
electroweak interactions, especially in order to clarify the 
geometrical meaning of curvatures in this geometry.
Considering the Higgs field to be a gauge field along $Z_2$ 
direction, we also discuss the BRST invariant gauge fixing 
in this theory.

\vskip 0.4cm\noindent
PACS number(s): 
\hfil
\vfill
\newpage}
\renewcommand{\theequation}{\thesection.\arabic{equation}}
\setcounter{equation}{0}
\addtocounter{section}{1}
\section*{\S\thesection.\ \ Introduction}

\indent

  The noncommutative geometry(NCG) of Connes\cite{Connes1,Connes2} 
has been successful in giving a geometrical interpretation of the 
standard model as well as some grand unification models.  
In this interpretation the Higgs fields are regarded as gauge fields 
along directions in the discrete space.  
The bosonic parts of actions are just the pure Yang-Mills actions 
containing gauge fields on both continuous and discrete spaces, 
and the Yukawa coupling is regarded as a kind of gauge interactions 
of fermions.

  There are now various alternative versions of NCG\cite{NCG}.  
Any NCG, however, has so far been algebraic rather than geometric.  
Nobody has considered enough the original geometric meaning such as 
covariant differences, parallel transportations, curvature and so on 
in the discrete space.
In previous works, one of the authors (T.S.) collaborated with 
Konisi has considered such a geometric meaning of NCG and proposed 
the gauge theory on $M_4\times Z_N$ without recourse to any 
knowledge of NCG, where $M_4$ is the four-dimensional Minkowski 
space and $Z_N$ is the discrete space with $N$ points.  
Here the Higgs fields have been introduced as mapping functions 
between any pair of vector fields belonging independently to the 
$N$-sheeted space-time, just as the Yang-Mills field is so between 
both vectors on $x$ and $x+\delta x$.  
We have applied this gauge theory to the Weinberg-Salam(WS) model 
for electroweak interactions, $N=2$ and $4$ super Yang-Mills 
theories and the Brans-Dicke theory for gravity\cite{Konisi}.

  In the present paper we revisit the above gauge theory 
on $M_4\times Z_N$ geometry, especially for the $Z_2$ case, 
because this has not so far been discussed enough.  
The WS model can be interpreted as the gauge theory on $M_4\times 
Z_2$ geometry, where the Higgs field is the gauge field associated 
with $Z_2$. 
In this geometry one can consider two kinds of curvatures for the 
Higgs field.
Our purpose is to clarify the geometrical meaning of these curvatures.  
We also discuss the BRST invariant gauge fixing in this gauge theory.  
At first sight one may wonder whether the Higgs field requires a new 
ghost, because it is a gauge field.  
This question becomes clear and is eventually reduced to the 
conventional $R_\xi$-gauge fixing plus Faddeev-Popov(FP) ghosts for 
the WS model with spontaneously broken symmetry of $SU(2)\times U(1)$.
There are now two similar works in this gauge fixing\cite{Lee}.  
Comparing with their works our approach which does not use NCG seems 
to be much simpler and clearer.

  In \S 2 we consider the gauge theory of the WS model on 
$M_4\times Z_2$, especially clarifying the geometrical meaning of 
curvatures for the Higgs field.
In \S 3 we discuss the BRST invariant gauge fixing plus FP ghosts  
in this geometry.  
The final section is devoted to concluding remarks.

\setcounter{equation}{0}
\addtocounter{section}{1}
\section*{\S\thesection.\ \ The Weinberg-Salam model on $M_4 \times 
Z_2$ geometry}

\indent

  To every point $(x,p)$ with $x\in M_4$ and $p\in Z_2$ we attach a 
complex $n_p$-dimensional internal vector space $V[n_p,x,p]$.
Let us take
\begin{eqnarray}
V[2,x,+]&=&{\rm complex\ 2\ dimensional\ space},\nonumber\\
V[1,x,-]&=&{\rm complex\ 1\ dimensional\ space},
\label{e201}\\
p&=&(+,-)\in Z_2.\nonumber
\end{eqnarray}

\noindent
The fermionic fields $\psi(x,p)$ are chosen as
\begin{eqnarray}
\psi(x,+)=\left(\matrix{\nu_L\cr e_L\cr}\right)
 \ \ \ {\rm and} \ \ \ \psi(x,-)=e_R,
\label{e202}
\end{eqnarray}

\noindent
where the left-handed neutrino $\nu_L$ and electron $e_L$ are the 
$SU(2)$ doublet and the right-handed electron $e_R$ is the $SU(2)$ 
singlet. 
Since $\nu_L$ and $e_L$  have the hypercharge $Y=-1$ and $e_R$ has 
$Y=-2$, the gauge field $\omega_\mu (x,p)$ coupled to $\psi(x,+)$ 
and $\psi(x,-)$ should be introduced as
\begin{eqnarray}
(\omega_\mu(x,+))^i_{\,j}
  &=&\frac{1}{2}[g'\tau^0 B_\mu(x)-g\tau^a A_\mu^a(x)]^i_{\,j},\ \ 
  (i,j=1,2)
\label{e203}\\
(\omega_\mu(x,-))^0_{\,0}
  &=&g'B_\mu(x),
\label{e204}
\end{eqnarray}

\noindent
respectively, where $\tau^a(a=1,2,3)$ is the Pauli matrix and 
$\tau^0$ is the $2\times 2$ unit matrix.  
The field strengths (or curvatures) for $\omega_\mu (x,\pm)$ are, 
therefore, given by
\begin{eqnarray}
F_{\mu\nu}(+)
  &=&\frac{1}{2}(g'\tau^0 B_{\mu\nu}-g\tau^a F_{\mu\nu}^a),
\label{e205}\\
F_{\mu\nu}(-)&=&g'B_{\mu\nu},
\label{e206}
\end{eqnarray}

\noindent
with
\begin{eqnarray}
B_{\mu\nu}&=&\partial_\mu B_\nu-\partial_\nu B_\mu,
\label{e207}\\
F_{\mu\nu}^a
  &=&\partial_\mu A_\nu^a-\partial_\nu A_\mu^a
    +g\epsilon_{abc}A_\mu^b A_\nu^c.
\label{e208}
\end{eqnarray}

  Let $\psi^i(x,p)$ be any vector field on $V[n_p,x,p]$.  We define 
covariant differences along the $Z_2$-direction by
\begin{eqnarray}
\delta_H\psi^i(x,+)
  &=&\psi^i(x,+)-H^i_{\,0}(x,+,-)\psi^0(x,-),
\label{e209}\\
\delta_H\psi^0(x,-)
  &=&\psi^0(x,-)-H^0_{\,i}(x,-,+)\psi^i(x,+),\ \ (i=1,2)
\label{e210}
\end{eqnarray}

\noindent
where $H^i_{\,0}(x,+,-)$ is the mapping function of $\psi^0(x,-)$ 
from $(x,-)$ to $(x,+)$ and $H^0_{\,i}(x,-,+)$ is that of 
$\psi^i(x,+)$ from $(x,+)$ to $(x,-)$.  
They are subject to transformation rules
\begin{eqnarray}
H^i_{\,0}(x,+,-)\rightarrow\widetilde H^i_{\,0}(x,+,-)
           =U^i_{\,j}(x,+)H^j_{\,0}(x,+,-)(U^{-1}(x,-))^0_{\,0},
\label{e211}\\
H^0_{\,i}(x,-,+)\rightarrow\widetilde H^0_{\,i}(x,-,+)
           =U^0_{\,0}(x,-)H^0_{\,j}(x,-,+)(U^{-1}(x,+))^j_{\,i},
\label{e212}
\end{eqnarray}

\noindent
under gauge transformations
\begin{eqnarray}
\psi^i(x,+)\rightarrow \tilde{\psi}^i(x,+)=U^i_{\,j}(x,+)\psi^j(x,+),
\label{e213}\\
\psi^0(x,-)\rightarrow \tilde{\psi}^0(x,-)=U^0_{\,0}(x,-)\psi^0(x,-).
\label{e214}
\end{eqnarray}

\noindent
where $U(x,\pm)$ are parametrized as
\begin{eqnarray}
U(x,\pm)=\exp\{i \theta(x,\pm)\}
\label{e215}
\end{eqnarray}

\noindent
with 
\begin{eqnarray}
\theta(x,+)=\frac{g'}{2}\tau^0 \theta(x)
            -\frac{g}{2}\tau^a \theta^a(x),\ \ 
\theta(x,-)=g' \theta(x).
\label{e216}
\end{eqnarray}

\noindent
Namely, $U(x,+)$ and $U(x,-)$ stand for local rotations of 
$SU(2)\times U_{\rm Y}(1)$ and $U_{\rm Y}(1)$, respectively.  
The rules (\ref{e211})--(\ref{e214}) guarantee vector properties of 
$\delta_H\psi^i(x,+)$ and $\delta_H\psi^0(x,-)$ on $V[n_p,x,p]$.  
The situation is quite the same as before for gauge fields 
$\omega_\mu(x,p)$, where the mapping functions are given by
\begin{eqnarray}
H(x+\delta x,x,p)=1+i\,\omega_\mu(x,p)\delta x^\mu+\cdots.
\label{e217}
\end{eqnarray}

\noindent
In this case a covariant variation of $\psi^i(x,p)$ due to 
an infinitesimal displacement $\delta x^\mu$ is given by
\begin{equation}
\delta\psi^i(x,p)=\psi^i(x+\delta x,p)
                  -\psi^i_{\parallel}(x+\delta x,p),
\label{e218}
\end{equation}

\noindent
where $\psi^i_{\parallel}(x+\delta x,p)$ is a 
parallel-transported vector of $\psi^i(x,p)$ from $x$ to 
$x+\delta x$, {\it i.e.},
\begin{equation}
\psi^i_{\parallel}(x+\delta x,p)=H^i_{\,j}(x+\delta x,x,p)\psi^j(x,p).
\label{e219}
\end{equation}

\noindent
Here, the mapping function $H^i_{\,j}$ should be subject to the 
transformation rule under the rotation $U(x,p)$
\begin{equation}
H(x+\delta x,x,p)\rightarrow\widetilde H(x+\delta x,x,p)
  =U(x+\delta x,p)H(x+\delta x,x,p)U^{-1}(x,p).
\label{e220}
\end{equation}

\noindent
This rule guarantees the vector property of 
$\psi_{\parallel}(x+\delta x,p)$ on $V[n_p,x+\delta x,p]$.  
If we use the familiar notations (\ref{e217}) and
\begin{eqnarray}
U(x+\delta x,p)&=&U(x,p)+\partial_\mu U(x,p)\delta x^\mu+\cdots
\label{e221}
\end{eqnarray}

\noindent
and keep terms up to the first-order $\delta x^\mu$ in 
Eq.(\ref{e220}), we have the gauge transformation rule for the 
non-Abelian gauge field $\omega_\mu(x,p)$
\begin{equation}
\omega_\mu(x,p)\rightarrow
  \tilde\omega_\mu(x,p)=U(x,p)\,\omega_\mu(x,p)U^{-1}(x,p)
  -i\partial_\mu U(x,p)U^{-1}(x,p).
\label{e222}
\end{equation}

  Comparing (\ref{e209}) -- (\ref{e212}) with (\ref{e218}) -- 
(\ref{e220})  we find that the mapping function $H(x,p,p')$ 
can be regarded as gauge fields associated with $Z_2$.  
Note that the gauge transformation rules (\ref{e211}) and 
(\ref{e212}) cannot be reduced to the form (\ref{e222}) with 
inhomogeneous terms, because $H^i_{\,0}(x,+,-)$ and 
$H^0_{\,i}(x,-,+)$ do not have Kronecker delta terms.
In order to guarantee the hermiticity of the 
Yukawa coupling terms with fermions which are nothing but gauge 
interactions, we assume
\begin{equation}
H^i_{\,0}(x,+,-)=(H^0_{\,i}(x,-,+))^*=H^i(x)=(H^+,H^0),
\label{e223}
\end{equation}

\noindent
where $*$ denotes the complex conjugation.
\ \\
\newlength{\minitwocolumn}
\setlength{\minitwocolumn}{.5\textwidth}
\addtolength{\minitwocolumn}{-.5\columnsep}
\begin{minipage}[t]{\minitwocolumn}
\hspace{5mm}
We now consider field strengths or curvatures for the gauge field 
$H(x,p,p')$.  
Let us call the usual field strength $F_{\mu\nu}(x,p)$ 
for $\omega_\mu(x,p)$ the curvature of the first type.
As well known, this comes from a difference between two parallel 
transportations of $\psi^i(x,p)$ along two paths depicted in Fig.1.  
\end{minipage}
\hspace{\columnsep}
\begin{minipage}[t]{\minitwocolumn}
\begin{center}
\ \vspace{-1ex}\\
\setlength{\unitlength}{1mm}
\begin{picture}(165,40)(10,-10)
 \put(50,10){\vector(-1,1){6}}\put(44,16){\line(-1,1){4}}
 \put(40,20){\vector(1,1){6}}\put(46,26){\line(1,1){4}}
 \put(50,10){\vector(1,1){6}}\put(56,16){\line(1,1){4}}
 \put(60,20){\vector(-1,1){6}}\put(54,26){\line(-1,1){4}}
 \put(50,10){\circle*{1}}
 \put(40,20){\circle*{1}}
 \put(60,20){\circle*{1}}
 \put(50,30){\circle*{1}}
 \put(39,32){$x+\delta_1 x+\delta_2 x$}
 \put(25,19){$x+\delta_1 x$}
 \put(62,19){$x+\delta_2 x$}
 \put(49,06){$x$}
 \put(25,06){{\bf Fig.1.}}
\end{picture}
\vspace{-1ex}\\
\end{center}
\end{minipage}
\begin{minipage}[t]{\minitwocolumn}
\hspace{5mm}
In quite the same way we consider a curvature $F_{\mu H}(x,+)$ 
of the second type which is defined by a difference between two 
mappings of $\psi^0(x,-)$ along paths $C_1$ and $C_2$ depicted 
in Fig.2.
The two mappings are given by
\end{minipage}
\hspace{\columnsep}
\begin{minipage}[t]{\minitwocolumn}
\begin{center}
\ \vspace{-1ex}\\
\setlength{\unitlength}{1mm}
\begin{picture}(165,40)(0,-15)
 \put(30,10){\line(1,0){10}}
 \put(30,25){\line(1,0){10}}
 \put(55,10){\line(1,0){10}}
 \put(55,25){\line(1,0){10}}
 \put(40,10){\vector(0,1){8}}\put(40,18){\line(0,1){7}}
 \put(40,25){\vector(1,0){8}}\put(48,25){\line(1,0){7}}
 \put(40,10){\vector(1,0){8}}\put(48,10){\line(1,0){7}}
 \put(55,10){\vector(0,1){8}}\put(55,18){\line(0,1){7}}
 \put(40,10){\circle*{1}}
 \put(40,25){\circle*{1}}
 \put(55,10){\circle*{1}}
 \put(55,25){\circle*{1}}
 \put(50,11){$C_1$}
 \put(41,21){$C_2$}
 \put(39,6){$x$}
 \put(50,6){$x+\delta x$}
 \put(39,27){$x$}
 \put(50,27){$x+\delta x$}
 \put(70,09){$-$}
 \put(70,24){$+$}
 \put(15,06){{\bf Fig.2.}}
\end{picture}
\vspace{-1ex}\\
\end{center}
\end{minipage}
\vspace{-2ex}
\begin{eqnarray}
C_1&=&H^i_{\,0}(x+\delta x,+,-)H^0_{\,0}(x+\delta x,x,-)\psi^0(x,-),
\label{e224}\\
C_2&=&H^i_{\,j}(x+\delta x,x,+)H^j_{\,0}(x,+,-)\psi^0(x,-).
\label{e225}
\end{eqnarray}

\noindent
Substituting (\ref{e217}) and (\ref{e223}) into $C_1$ and $C_2$ above, 
we have 
\begin{eqnarray}
C_1-C_2&=&[\partial_\mu H(+,-)
                   +i\,\omega_\mu(+)H(+,-)
                   -iH(+,-)\,\omega_\mu(-)]\psi^0(-)\delta x^\mu
\nonumber\\
  &=&[\partial_\mu H(x)-\frac{i}{2}(g'\tau^0 B_\mu
                          +g\tau^a A_\mu^a)H(x)]\psi^0(-)\delta x^\mu
\nonumber\\
  &\equiv&(\nabla_{\!\mu}H(x))\psi^0(-)\delta x^\mu.
\label{e227}
\end{eqnarray}

\noindent
The second type curvature components $F_{\mu H}(x,+)$ are, 
therefore, given by 
\begin{eqnarray}
(F_{\mu H}(x,+))^i_{\,0}=(\nabla_{\!\mu}H(x))^i.
\label{e228}
\end{eqnarray}

\noindent
In the same way we have
\begin{eqnarray}
(F_{\mu H}(x,-))^0_{\,i}=(\nabla_{\!\mu}H(x))^{\dagger}_i.
\label{e229}
\end{eqnarray}
\begin{minipage}[t]{\minitwocolumn}
\hspace{5mm}
  A curvature of the third type \break $F_{HH}(x,+)$ corresponds to 
Fig.3.  
Namely, $\psi^i(x,+)$ is compared with $\psi^i_\parallel(x,+)$ which 
is the mapped function of $\psi^i(x,+)$ from $(x,+)$ to $(x,-)$ and 
then returning to $(x,+)$, {\it i.e.}, 
\end{minipage}
\hspace{\columnsep}
\begin{minipage}[t]{\minitwocolumn}
\begin{center}
\ \vspace{-1ex}\\
\setlength{\unitlength}{1mm}
\begin{picture}(125,30)(-10,-5)
 \put(25,5){\line(1,0){10}}
 \put(25,20){\line(1,0){10}}
 \put(35,5){\line(1,0){10}}
 \put(35,20){\line(1,0){10}}
 \put(36,5){\vector(0,1){8}}\put(36,13){\line(0,1){7}}
 \put(35,20){\vector(0,-1){8}}\put(35,12){\line(0,-1){7}}
 \put(35.5,5){\circle*{1}}
 \put(35.5,20){\circle*{1}}
 \put(30,1){$(x,+)$}
 \put(30,22){$(x,-)$}
 \put(10,1){{\bf Fig.3.}}
\end{picture}
\vspace{-1ex}\\
\end{center}
\end{minipage}
\begin{eqnarray}
\psi^i(x,+)-\psi^i_\parallel(x,+)
  &=&\psi^i(x,+)-H^i_{\,0}(x,+,-)H^0_{\,j}(x,-,+)\psi^j(x,+)
\nonumber\\
  &=&(\delta^i_{\,j}-H^i(x)H^\dagger_j(x))\psi^j(x,+).
\label{e230}
\end{eqnarray}

\noindent
This gives the third type curvature 
\begin{eqnarray}
(F_{HH}(x,+))^i_{\,j}=\delta^i_{\,j}-H^i(x)H^\dagger_j(x).
\label{e231}
\end{eqnarray}

\noindent
In the same way we have 
\begin{eqnarray}
(F_{HH}(x,-))^0_{\,0}=1-H^\dagger_i(x)H^i(x).
\label{e232}
\end{eqnarray}

\noindent
On $M_4$ we know that there is no curvature of the similar type 
corresponding to two paths: $x\rightarrow x+\delta x\rightarrow x$ 
and $x\rightarrow x$.  
On $Z_2$, however, we find non-vanishing curvature of the third type 
(see Appendix).

  Now, considering $A^a_\mu(x)$, $B_\mu(x)$ and $H^i(x)$ to be 
gauge fields, a Lagrangian for them should be of the Yang-Mills type 
\begin{equation}
{\cal L}={\cal L}_1+{\cal L}_2+{\cal L}_3,
\label{e233}
\end{equation}

\noindent
where
{\addtocounter{equation}{-1}
\setcounter{enumi}{\value{equation}}
\addtocounter{enumi}{1}
\setcounter{equation}{0}
\renewcommand{\theequation}{\thesection.\theenumi\alph{equation}}
\begin{eqnarray}
{\cal L}_1&=&-\displaystyle{\sum_p}\frac{1}{4c_p^2}\,{\rm tr}
            [F^\dagger_{\mu\nu}(p)F^{\mu\nu}(p)]\nonumber\\
          &=&-\frac{1}{8c_+^2}(g^2 F_{\mu\nu}^a F^{a\mu\nu}
                              +{g'}^2 B_{\mu\nu} B^{\mu\nu})
             -\frac{1}{4c_-^2}{g'}^2 B_{\mu\nu} B^{\mu\nu}\nonumber\\
          &=&-\frac{g^2}{8c_+^2}F_{\mu\nu}^a F^{a\mu\nu}
             -(\frac{{g'}^2}{8c_+^2}+\frac{{g'}^2}{4c_-^2})
               B_{\mu\nu} B^{\mu\nu},
\label{e233a}\\
{\cal L}_2&=&-\displaystyle{\sum_p}\xi_p\,{\rm tr}
            [F_{\mu H}^\dagger(p)F^{\mu H}(p)]\nonumber\\
          &=&-(\xi_++\xi_-)(\nabla_{\!\mu}H)^{\dagger}(\nabla^\mu H),
\label{e233b}\\
{\cal L}_3&=&-\displaystyle{\sum_p}\zeta_p\,{\rm tr}
            [F_{HH}^\dagger(p)F^{HH}(p)]\nonumber\\
          &=&-\zeta_+\,{\rm tr}(1-HH^\dagger)(1-HH^\dagger)
             -\zeta_-(1-H^\dagger H)^2\nonumber\\
          &=&-(\zeta_++\zeta_-)\biggl[(H^\dagger H-1)^2
             +\frac{\zeta_+}{\zeta_++\zeta_-}\biggr],
\label{e233c}
\end{eqnarray}
\setcounter{equation}{\value{enumi}}
}

\noindent
where $c_p$, $\xi_p$ and $\zeta_p$ are normalization constants.  
They should be so chosen as to be consistent with positivity of 
kinetic terms and renormalizability of the theory.
Let us normalize ${\cal L}_1$ by
\begin{equation}
{\cal L}_1=-\frac{1}{4}(F_{\mu\nu}^a F^{a\mu\nu}
                                       +B_{\mu\nu} B^{\mu\nu}),
\label{e234}
\end{equation}

\noindent
so that
\begin{equation}
g=\sqrt{2}c_+,\ \ g'=\frac{\sqrt{2}c_+ c_-}{\sqrt{2c_+^2+c_-^2}}.
\label{e235}
\end{equation}

\noindent
This means that $g$ and $g'$ are still independent parameters with 
each other.  
If we redefine the scalar field $H$ by
\begin{equation}
\phi=\sqrt{-\xi_+ -\xi_-}\,H,
\label{e236}
\end{equation}

\noindent
${\cal L}_2$ and ${\cal L}_3$ are reduced to the original WS type
\begin{eqnarray}
{\cal L}_2&=&(\nabla_{\!\mu}\phi)^\dagger(\nabla^\mu\phi),
\label{e237}\\
{\cal L}_3&=&-\mu^2|\phi|^2-\lambda|\phi|^4+{\rm const.},
\label{e238}
\end{eqnarray}

\noindent
where
\begin{equation}
\lambda=\frac{\zeta_++\zeta_-}{(\xi_++\xi_-)^2},\ \  
  \mu^2=2\frac{\zeta_++\zeta_-}{\xi_++\xi_-}=2(\xi_++\xi_-)\lambda<0. 
\label{e239}
\end{equation}

\noindent
Here we have assumed to be $\xi_++\xi_-<0$ and $\zeta_++\zeta_->0$.
The ${\cal L}_3$ is nothing but the Higgs potential which has a 
minimal value at $H^{\dagger}H=1$, {\it i.e.}, 
$|\phi|^2=-(\xi_++\xi_-)=-\mu^2/2\lambda>0.$  
Both constants $\lambda$ and $\mu^2$ are still independent 
parameters with each other within $\lambda>0$ and $\mu^2<0$.
The fermionic part will be neglected because it is irrelevant to 
our purpose, it can be seen in Ref.4.

\setcounter{equation}{0}
\addtocounter{section}{1}
\section*{\S\thesection.\ \ Gauge fixing and FP ghosts}

\indent

  In \S 2  we have seen that the WS model can be interpreted as the 
gauge theory on $M_4\times Z_2$ geometry, where the Higgs field is 
the gauge field along the direction in the discrete space $Z_2$.  
The Higgs fields obeys the gauge transformation rule (\ref{e211}), 
{\it i.e.}, 
\begin{equation}
H(+,-)\rightarrow\widetilde H(+,-)=U(+)H(+,-)U^{-1}(-).
\label{e301}
\end{equation}

\noindent
For an infinitesimal gauge transformation 
$U(\pm)\cong 1+i\theta(\pm)$, Eq.(\ref{e301}) becomes in the 
notations (\ref{e236}) and (\ref{e216}) 
\begin{eqnarray}
\delta\phi&=&i\,\theta(+)\phi-i\,\phi\theta(-) \nonumber\\
          &=&-\frac{i}{2}(g'\tau^0\theta+g\tau^a\theta^a)\phi.
\label{e302}
\end{eqnarray}

\noindent
This shows that the $\phi$ has the hypercharge $Y=+1$ and coincides 
with the conventional Higgs scalar field.
The BRST transformation $\delta_B\phi$ is then obtained by replaceing 
$\theta$ and $\theta^a$ by ghosts $c^0$ and $c^a$
\begin{eqnarray}
\delta_B\phi=-\frac{i}{2}(g'\tau^0 c^0+g\tau^a c^a)\phi.
\label{e303}
\end{eqnarray}

\noindent
For other gauge fields $B_\mu$ and $A^a_\mu$ it follows from 
(\ref{e222}) that their BRST transformations are given by 
\begin{eqnarray}
\delta_B B_\mu=\partial_\mu c^0\ \ {\rm and}\ \ 
\delta_B A^a_\mu=\partial_\mu c^a+g\epsilon_{abc}A^b c^c
\equiv D_\mu c^a.
\label{e304}
\end{eqnarray}

\noindent
Eq.(\ref{e303}) shows that the $\phi$ does not require any new 
ghosts though it is a gauge field.

  Thus we have seen that the gauge-fixing in our geometry is reduced 
to the conventional one.  
For completeness we give the full result of the BRST invariant 
$R_\xi$-gauge fixing plus FP ghosts with spontaneously broken 
symmetry of $SU(2)\times U_{\rm Y}(1)$.  
The BRST invariant Lagrangian for gauge fixing plus FP ghosts 
should be of the form\cite{Kugo} 
\begin{equation}
{\cal L}_{\rm GF+FP}=-i\delta_{\rm B}(*),
\label{e305}
\end{equation}

\noindent
where $*$ is chosen as follows:
\begin{eqnarray}
*&=&\bar c^0
  (\partial^\mu B_\mu+\frac{1}{2}\alpha B^0+\alpha M_W \chi^0)
\nonumber\\
 &&+\bar c^a
  (\partial^\mu A^a_\mu+\frac{1}{2}\alpha B^a+\alpha M_W \chi^a).
\label{e306}
\end{eqnarray}

\noindent
Here, $\bar c^0$ and $\bar c^a$ are anti-ghosts, 
Nakanishi-Lautrup fields $B^0$ and $B^a$ are defined by 
$-i\delta_{\rm B}\bar c^0=B^0$ and 
$-i\delta_{\rm B}\bar c^a=B^a$, $\alpha$ the gauge parameter, 
and $M_W$ a mass parameter.  
The Higgs field $\phi$ is parametrized as 
\begin{equation}
\phi=\left(\matrix{\phi^+\cr \phi^0\cr}\right)
    =\frac{1}{\sqrt{2}}\left(\matrix{\chi^2(x)+i\chi^1(x)\cr
                                     v+\psi(x)-i\chi^3(x)\cr}\right),
\label{e307}
\end{equation}

\noindent
where $\psi(x),\chi^a(x),a=1,2,3,$ are real scalar fields and $v$ 
is a real constant.
The $\chi^a$ has been used in (\ref{e306}),while the $\chi^0$ 
is the $U_{\rm Y}(1)$ phase factor of $\phi$.

  Now, ${\cal L}_{\rm GF+FP}$ turns out to be of the form
\begin{equation}
{\cal L}_{\rm GF+FP}={\cal L}_{\rm GF}+{\cal L}_{\rm FP},
\end{equation}

\noindent
where
\begin{eqnarray}
{\cal L}_{\rm GF}&=&B^0
  (\partial^\mu B_\mu+\frac{1}{2}\alpha B^0+\alpha M_W \chi^0)
\nonumber\\
  &&+B^a
  (\partial^\mu A^a_\mu+\frac{1}{2}\alpha B^a+\alpha M_W \chi^a),
\label{e309} \\
{\cal L}_{\rm FP}&=&i\,\bar c^0
  (\partial^\mu\delta_{\rm B}B_\mu+\alpha M_W\delta_{\rm B}\chi^0)
\nonumber\\
  &&+i\,\bar c^a
  (\partial^\mu\delta_{\rm B}A^a_\mu+\alpha M_W\delta_{\rm B}\chi^a).
\label{e310}
\end{eqnarray}

\noindent
After rotating $(B_\mu,A_\mu^3)$, $(B^0,B^3)$ and $(\chi^0,\chi^3)$ 
to $(A_\mu,Z_\mu)$, $(B_A,B_Z)$ and $(\chi_A,\chi_Z)$ by the 
Weinberg angle $\theta_W$, respectively, {\it i.e.},
\begin{eqnarray}
&&A_\mu=\cos\theta_W B_\mu+\sin\theta_W A_\mu^3,\ \ 
B_A=\cos\theta_W B^0+\sin\theta_W B^3,\nonumber\\
&&Z_\mu=-\sin\theta_W B_\mu+\cos\theta_W A_\mu^3,\ \ 
B_Z=-\sin\theta_W B^0+\cos\theta_W B^3,
\label{e311}
\end{eqnarray}

\noindent
and
\begin{eqnarray}
&&\chi_A=\cos\theta_W\chi^0+\sin\theta_W\chi^3,\nonumber\\
&&\chi_Z=-\sin\theta_W\chi^0+\cos\theta_W\chi^3,
\label{e312}
\end{eqnarray}

\noindent
we have
\begin{eqnarray}
{\cal L}_{\rm GF}
  &=&B_Z(\partial^\mu Z_\mu+\frac{1}{2}\alpha B_Z+\alpha M_W \chi_Z)
    +B_A(\partial^\mu A_\mu+\frac{1}{2}\alpha B_A+\alpha M_W \chi_A)
\nonumber\\
  &&+B^1(\partial^\mu A_\mu^1+\frac{1}{2}\alpha B^1+\alpha M_W \chi^1)
   +B^2(\partial^\mu A_\mu^2+\frac{1}{2}\alpha B^2+\alpha M_W \chi^2).
\label{e313}
\end{eqnarray}

\noindent
If we set $\chi_A=0$, then it follows
\begin{equation}
\chi^0=-\tan\theta_W\chi^3=-\frac{g'}{g}\chi^3,
\label{e314}
\end{equation}

\noindent
hence
\begin{equation}
\chi_Z=\frac{1}{\cos\theta_W}\chi^3=\frac{M_Z}{M_W}\chi^3.
\label{e315}
\end{equation}

\noindent
Thus, finally we obtain the $R_\xi$-gauge fixing 
Lagrangian\cite{Fujikawa}
\begin{eqnarray}
{\cal L}_{\rm GF}
  &=&\frac{1}{2}\alpha(B_1^2+B_2^2+B_Z^2+B_A^2)
    +B_1(\partial^\mu A_\mu^1+\alpha M_W\chi^1)
    +B_2(\partial^\mu A_\mu^2+\alpha M_W\chi^2)\nonumber\\
  &&+B_Z(\partial^\mu Z_\mu+\alpha M_Z\chi^3)
    +B_A\partial^\mu A_\mu.
\label{e316}
\end{eqnarray}

\noindent
Eliminating $B$-fields from ${\cal L}_{\rm GF}$, it is reduced to
\begin{equation}
{\cal L}_{\rm GF}=-\frac{1}{2\alpha}\displaystyle\sum_{a=1}^2
    (\partial^\mu A_\mu^a+\alpha M_W\chi^a)^2
    -\frac{1}{2\alpha}(\partial^\mu Z_\mu+\alpha M_Z\chi^3)^2
    -\frac{1}{2\alpha}(\partial^\mu A_\mu)^2.
\label{e317}
\end{equation}

\noindent
In order to obtain the FP-ghost Lagrangian ${\cal L}_{\rm FP}$, 
we need $\delta_{\rm B}\chi^a$ in (\ref{e310}).
Substituting the parametrization (\ref{e307}) of $\phi$ into 
(\ref{e303}), we have
\begin{eqnarray}
\delta_{\rm B}\chi^1&=&-\frac{g}{2}(\chi^2 c^3-\chi^3 c^2)
  -\frac{g}{2}c^1(v+\psi)-\frac{g'}{2}c^0\chi^2,\nonumber\\
\delta_{\rm B}\chi^2&=&-\frac{g}{2}(\chi^3 c^1-\chi^1 c^3)
  -\frac{g}{2}c^2(v+\psi)+\frac{g'}{2}c^0\chi^1,\nonumber\\
\delta_{\rm B}\chi^3&=&-\frac{g}{2}(\chi^1 c^2-\chi^2 c^1)
  -\frac{g}{2}c^3(v+\psi)+\frac{g'}{2}c^0(v+\psi),\nonumber\\
\delta_{\rm B}\psi&=&\frac{g}{2}(\chi^1 c^1+\chi^2 c^2+\chi^3 c^3)
  -\frac{g'}{2}c^0\chi^3.
\label{e318}
\end{eqnarray}

\noindent
By using the relation $\chi^0=-(g'/g)\chi^3$ and Eq.(\ref{e304}),
the ghost Lagrangian ${\cal L}_{\rm FP}$ becomes
\begin{eqnarray}
{\cal L}_{\rm FP}
  &=&-i\partial^\mu\bar c^0\partial_\mu c^0
     -i\partial^\mu\bar c^a D_\mu c^a
     -i\alpha M_W\frac{g}{2}
       [\bar c^a(\vec\chi\times\vec c)^a+\bar c^a c^a(v+\psi)]
\nonumber\\
   &&-i\alpha M_W\frac{g'}{2}
       [\bar c^0(c^1\chi^2-c^2\chi^1)
        +(\bar c^1\chi^2-\bar c^2\chi^1)c^0
        -(\bar c^3 c^0+\bar c^0 c^3)(v+\psi)]
\nonumber\\
   &&-i\alpha M_W\frac{g'}{2g}
       \bar c^0 c^0(v+\psi).
\label{e319}
\end{eqnarray}

\noindent
Thus we have obtained the BRST invariant $R_\xi$-gauge fixing and 
FP-ghost Lagrangians (\ref{e316}) and (\ref{e319}).  

\setcounter{equation}{0}
\addtocounter{section}{1}
\section*{\S\thesection.\ \ Concluding remarks}

\indent

  We have found that the covariant derivative of the Higgs field 
$\nabla_{\!\mu}H$ is just the curvature of the second type 
corresponding to Fig.2 and the term $1-HH^\dagger$ is that of the 
third type corresponding to Fig.3.  
Since the Higgs field is one of gauge fields, its Lagrangian 
should be of Yang-Mills type.  
We have seen that this Lagrangian coincides exactly with that of the 
WS type with the Higgs potential.

  We have also shown that the Higgs field does not require any 
new ghost though it is the gauge field.  
The BRST transformation of $H$ coincides with the conventional 
one.  
Then we have given the full and new result of the BRST invariant 
$R_\xi$-gauge fixing plus FP ghosts for the WS model with spontaneous 
broken symmetry of $SU(2)\times U_{\rm Y}(1)$.

\vspace{1.0cm}

\indent
{\large \bf Acknowledgments}:
  We thank G. Konisi and K. Shigemoto for useful discussions and 
invaluable comments.
Thanks are also due to Z. Maki for his interest in this work and 
encouraging us.

\newpage
\renewcommand{\theequation}{\Alph{section}.\arabic{equation}}
\setcounter{equation}{0}
\setcounter{section}{1}
\section*{Appendix}

\indent

  We would like to show that on $M_4$ there is no curvature of 
the third type in the text.  
Let us consider two paths on $M_4$
\begin{eqnarray}
C_1&:&\ \ x\rightarrow x+\delta_1 x
           \rightarrow (x+\delta_1 x)+\delta_2 x,
\label{eA01}\\
C_2&:&\ \ x\rightarrow x+(\delta_1 x+\delta_2 x).
\label{eA02}
\end{eqnarray}

\ \\
\begin{minipage}[t]{\minitwocolumn}
The difference $\Delta=C_1-C_2$ between two mapping functions is 
given by 
\end{minipage}
\hspace{\columnsep}
\begin{minipage}[t]{\minitwocolumn}
\begin{center}
\ \vspace{-1ex}\\
\setlength{\unitlength}{1mm}
\begin{picture}(125,30)(-10,-5)
 \put(30,5){\vector(0,1){11}}\put(30,16){\line(0,1){9}}
 \put(30,5){\vector(1,1){6}}\put(36,11){\line(1,1){4}}
 \put(40,15){\vector(-1,1){6}}\put(34,21){\line(-1,1){4}}
 \put(30,5){\circle*{1}}
 \put(40,15){\circle*{1}}
 \put(30,25){\circle*{1}}
 \put(24,14){$C_1$}
 \put(34,14){$C_2$}
 \put(19,27){$x+\delta_1 x+\delta_2 x$}
 \put(42,14){$x+\delta_1 x$}
 \put(29,1){$x$}
 \put(5,1){{\bf Fig.4.}}
\end{picture}
\vspace{-1ex}\\
\end{center}
\end{minipage}
\begin{eqnarray}
\Delta&=&C_1-C_2 \nonumber\\
      &=&H((x+\delta_1 x)+\delta_2 x,x+\delta_1 x)
       H(x+\delta_1 x,x)-H(x+(\delta_1 x+\delta_2 x),x).
\label{eA03}
\end{eqnarray}

\noindent
Substituting the expression of $H$
\begin{eqnarray}
H(x+\delta x,x)=1+i\,\omega_\mu(x)\delta x^\mu
                 +\frac{i}{2}C_{\mu\nu}(x)\delta x^\mu\delta x^\nu
                 +\cdots, 
\label{eA04}
\end{eqnarray}

\noindent
where $C_{\mu\nu}$ is symmetric with respect to $\mu$ $\nu$ 
interchanged, into (\ref{eA03}), we have 
\begin{eqnarray}
\Delta=(-\omega_\mu\omega_\nu+i\,\omega_{\nu,\mu}
        -i\,C_{\mu\nu})\delta_1 x^\mu\delta_2 x^\nu.
\label{eA05}
\end{eqnarray}

\noindent
If we choose $\delta_2 x^\mu=\alpha\delta_1 x^\mu\ (\alpha>0)$, two 
paths $C_1$ and $C_2$ become the same.  
In this case the difference $\Delta$ vanishes so that 
\begin{eqnarray}
C_{\mu\nu}=\frac{1}{2}(i\,\omega_{\{\mu}\omega_{\nu\}}
                                 -\omega_{\{\nu,\mu\}}).
\label{eA06}
\end{eqnarray}

\noindent
Substituting this into (\ref{eA05}) we find 
\begin{eqnarray}
\Delta&=&-\frac{i}{2}\{\partial_\mu\omega_\nu-\partial_\nu\omega_\mu
                  +i\,[\omega_\mu,\omega_\nu]\}
                   \delta_1 x^\mu\delta_2 x^\nu\nonumber\\
      &=&-\frac{i}{2}F_{\mu\nu}\delta_1 x^\mu\delta_2 x^\nu,
\label{eA07}
\end{eqnarray}

\noindent
which corresponds to one half of curvature for $\omega_\mu$.

  Let us now choose $\alpha=-1$, {\it i.e.}, 
$\delta_2 x^\mu=-\delta_1 x^\mu=-\delta x^\mu$.  
Then Eq.(\ref{eA03}) is reduced to 
\begin{eqnarray}
\Delta&=&H(x,x+\delta x)H(x+\delta x,x)-1 \nonumber\\
      &=&\frac{i}{2}F_{\mu\nu}\delta x^\mu\delta x^\nu \nonumber\\
      &=&0,
\label{eA08}
\end{eqnarray}

\noindent
because $F_{\mu\nu}$ is antisymmetric with respect to $\mu$ and $\nu$.  
This shows that on $M_4$ there is no curvature of the third type 
corresponding to two paths: $x\rightarrow x+\delta x\rightarrow x$ 
and $x\rightarrow x$.

  In the discrete space, say $Z_N$, there is no case such that two 
paths $A\rightarrow B\rightarrow C$ and $A\rightarrow C$ become 
the same.  
So we calculate directly $\Delta$ in $Z_2$ defined by 
\begin{eqnarray}
\Delta&=&1-H(+,-)H(-,+).
\label{eA09}
\end{eqnarray}

\noindent
This is nothing but the curvature of the third type (see Fig.3).


\end{document}